# How Work From Home Affects Collaboration: A Large-Scale Study of Information Workers in a Natural Experiment During COVID-19

Longqi Yang, Sonia Jaffe, David Holtz, Siddharth Suri, Shilpi Sinha, Jeffrey Weston, Connor Joyce, Neha Shah, Kevin Sherman, CJ Lee, Brent Hecht, Jaime Teevan

Microsoft Corporation
Correspondence: Longqi.Yang@microsoft.com

**ABSTRACT**

The COVID-19 pandemic has had a wide-ranging impact on information workers such as higher stress levels, increased workloads, new workstreams, and more caregiving responsibilities during lockdown. COVID-19 also caused the overwhelming majority of information workers to rapidly shift to working from home (WFH). The central question this work addresses is: can we isolate the effects of WFH on information workers' collaboration activities from all other factors, especially the other effects of COVID-19? This is important because in the future, WFH will likely to be more common than it was prior to the pandemic.

We use difference-in-differences (DiD), a causal identification strategy commonly used in the social sciences, to control for unobserved confounding factors and estimate the causal effect of WFH. Our analysis relies on measuring the difference in changes between those who WFH prior to COVID-19 and those who did not. Our preliminary results suggest that on average, people spent more time on collaboration in April (Post WFH mandate) than in February (Pre WFH mandate), but this is primarily due to factors other than WFH, such as lockdowns during the pandemic. The change attributable to WFH specifically is in the opposite direction: less time on collaboration and more focus time. This reversal shows the importance of using causal inference: a simple analysis would have resulted in the wrong conclusion. We further find that the effect of WFH is moderated by individual remote collaboration experience prior to WFH. Meanwhile, the medium for collaboration has also shifted due to WFH: instant messages were used more, whereas scheduled meetings were used less. We discuss design implications -- how future WFH may affect focused work, collaborative work, and creative work.

**KEYWORDS**

Work-from-home, remote work, COVID-19, causal inference, information worker

## 1 Introduction

The COVID-19 pandemic has affected information workers in myriad ways: for many school closures have given them additional caregiving responsibilities, lockdowns have restricted their nonwork activities, and health concerns have raised stress levels. The pandemic also forced many information workers to switch to work from home (WFH) to avoid catching the disease. Because of the confounding factors caused by the pandemic, one cannot infer work-from-home effects from pure observational changes measured before and after COVID-19. There may also be simultaneous factors unrelated to the pandemic that affected information workers, such as changes in workloads due to seasonality. Thus, the research question this work addresses is: can we isolate the effect of switch to working from home on information workers' collaboration activities from all other factors?

The effect of WFH is an important research question not only for supporting work during the COVID-19 crisis, but also for informing what will happen in the future. During COVID-19, 34.1% of Americans switched to work from home [5]. It is estimated that 37% of jobs in United States can be done remotely [7], and many technology companies have decided to embrace much more remote work moving forward [1,12,14,18]. This suggests that whatever equilibrium society reaches post-COVID-19, it will likely be one between the extremes of little WFH and complete WFH. The goal of our research is to inform such future using the almost overnight shift as a natural experiment.

In this paper, we conduct causal inference with a difference-in-differences (DiD) analysis of large-scale software usage logs from Microsoft to disentangle the causal effects of WFH from other confounding factors. Intuitively, our analysis works as follows. Any change in the behavior of those already working from home prior to COVID-19 was due to other factors beyond the WFH transition. On the other hand, any change in the behavior of those working in the office prior to COVID-19 was due to other factors combined with the move to working from home. If the trends were parallel to begin with, taking the difference allows us to isolate the effect of working from home. We also analyze how the effects are moderated by remote collaboration experience prior to WFH.

Our initial study characterizes people's collaboration behavior through the lens of focus hours, and the hours spent on scheduled meetings, messaging, and emails. We find that

- Though on average total collaboration hours were higher and focus hours were lower in April relative to February,[1] **WFH specifically caused an opposite change** – WFH decreased collaboration hours (-2.0%) and increased focus hours (+3.6%). The observational trends would have resulted in the opposite, wrong conclusion.
- WFH caused people to shift between collaboration tools: while email hours stayed stable, scheduled meeting hours were decreased (-6.4%), and messaging hours were increased (+27.7%). The increase in messaging due to WFH accounts for less than half of the observed increase, which further justifies the need of using causal inference to control for confounding factors.
- The effect differs by individuals' experience with remote collaboration; among those who previously worked in an office, we find that **the effects are primarily for those who had less remote collaboration experience** (i.e., less than 20% of collaborations were with co-workers in different cities or WFH), compared to those who previously did at least 70% of their collaboration remotely. This suggests that for most outcomes, the effect of WFH is due to workers' need to collaborate remotely, not to their physically working from home.

In total, our findings suggest that WFH could impact not only the usage of productivity tools, but also how workers accomplish different types of work. A shift to WFH may be beneficial for those engaging in focused work that requires large blocks of free time but may be detrimental for those engaging in work that is highly collaborative in nature. Our study also highlights the importance of causal inference. Simply comparing worker behavior before and during the pandemic does not accurately measure the effects of WFH, and in many cases it suggests an effect with the opposite sign of the actual effect. If we hope to use the COVID-19 pandemic as an occasion to predict the effects of future WFH, it is critical to separate the effect of WFH from other confounding factors. We plan to extend this research by investigating the impacts of WFH on a broader set of measures for both work practices and productivity.

## 2 Related Work

The majority of prior studies related to telecommuting and remote work are based on cross-sectional data (see survey papers [2,16] for details), which cannot be directly used to draw causal conclusions regarding the impact of WFH. There have been two large studies using experimental variation to analyze WFH effect. Bloom et al. [4] analyzed a travel company that randomly allowed half of its call center workers who were interested in working from home, and had sufficient broadband and space, to do so. They found that employees working from home missed fewer days of work and took less break time during the workday, leading to a 13% increase in calls answered. Choudhury et al. [6] analyzed a policy change for employees of the US Patent Office. Due to union negotiations, the policy allowing "work from anywhere" was rolled out in three phases, with employees randomly assigned to a phase. The study found a 4% increase in productivity.[2]

---
[1] The data was from the year of 2020.

Our study makes several contributions relative to prior causal work: (1) We look at changes in finer-grained aspects of people's work practice and behavior, such as time spent on collaboration, meetings etc. (2) Since WFH is mandatory for most employees during COVID-19, we measure the effect of WFH for the average office worker as opposed to the effect for those who volunteer or request WFH. (3) We study a broad population of information workers, making our results potentially more generalizable than evidence from a single very specific profession. Our findings can also inform better design of productivity tools for remote work [8,11,13,17,19], which is an important topic of research for Human-Computer Interaction (HCI) and Computer Supported Cooperative Work (CSCW).

## 3 Methodology

Prior to COVID-19, some employees worked from home, whereas others worked in an office. However, a simple comparison of these two groups is unlikely to provide an unbiased estimate of the causal effect of WFH, since employees who work from home are likely to differ from those who do not in important and potentially unobservable ways, e.g., the content of their jobs and/or their working styles. Similarly, a simple comparison of pre-COVID-19 and post-COVID-19 outcomes for employees who were previously working in an office is unlikely to generate credible causal estimates, since any observed differences may be driven by many factors that are unrelated to WFH, such as seasonal changes in work behavior and productivity or COVID-19-related impacts, e.g., childcare responsibilities, social isolation and existential dread. In order to obtain a credible causal estimate of the effect of WFH on collaboration behavior, we employ a difference-in-differences identification strategy, which takes into account both time invariant differences between the two groups of workers and temporal trends that are common to the two groups of workers.

### 3.1 Background: difference-in-differences

We use a difference-in-differences (DiD) identification framework [3] to separate the causal impact of WFH on collaboration metrics from differences between groups and over time. In a simple version of DiD where the researcher does not control for additional covariates, the difference-in-differences framework assumes that the outcome for person $i$ belonging to group $g$ at time $t$, $y_{igt}$, can be modeled as

$$y_{igt} = \gamma_g + \lambda_t + \delta \cdot d_{gt} + \varepsilon_{igt}, \quad (1)$$

where $\gamma_g$ is a group-level fixed effect that accounts for time-invariant differences between groups, $\lambda_t$ is a time fixed effect that accounts for any temporal trends that affect all groups in the same way, $d_{gt}$ indicates whether or not group $g$ has been treated at time $t$, and $\varepsilon_{igt}$ is a residual with $E(\varepsilon_{igt} | g,t) = 0$.

---

[2] It should be noted that employees were already allowed to work from home four days a week; while the effect could be due to employees not having to commute the one-day week, the authors attribute it mostly to employees moving out of the Washington DC area – and perhaps being able to afford a better home office or being more motivated.

The treatment effect, $\delta$, is causally identified so long as Eq. 1 accurately represents the data generating process. That is, there must be one time effect (e.g. COVID or seasonality), captured by $\lambda_t$, that applies to both the treatment and control group in each time period, so if, counter-factually, the treatment group had not been treated at time $t$, the expected average outcome for the group would have been $\gamma_g + \lambda_t$. Equivalently, the average expected change in outcome between periods $t$ and $t'$ in the counterfactual where the group is not treated would have been $\lambda_{t'} - \lambda_t$, the same as the control group. It is impossible to verify that this "parallel trends" assumption holds, given that the outcome absent treatment is not observed for people after they have been treated. However, the credibility of the parallel trends assumption is often established by verifying that pre-treatment time series trends are parallel [9].

## 3.2 Our difference-in-differences model

In our context, those who were working from home prior to COVID-19 ($g$ = P-WFH) are the "control group" for the natural experiment: the change in their outcomes during the WFH mandate captures the effects of seasonality, product cycles, social isolation, etc. The difference between the control group's change in outcome and the change for those who had previously worked in an office ($g$ = P-WFH) is attributed to the latter group's having switched to working from home. In other words, the treatment indicator $d_{gt}$ in Eq. 1 can be expressed as

$$d_{gt} = \mathbb{1}(g = \text{P-OFC}) \cdot \mathbb{1}(t = \text{post-COVID-19}), \quad (2)$$

where $\mathbb{1}(x) = 1$ if x is true. This means that in contrast to the classic difference-in-differences setup, our difference-in-differences model relies on two identifying assumptions. First, that the time series for the two groups would have moved in parallel absent COVID-19, and second, that the non-WFH effects of COVID-19 were on average the same for those who were previously working from home and those who were previously working in an office.

In order to make our identifying assumptions more credible, we use coarsened exact matching [10] to reweight control group observations, so that the control and treatment groups are balanced with respect to three covariates: job role, managerial status, and a binary measure of level (i.e., seniority). This makes our model robust to deviations from the "parallel trends" assumption that can be explained by these observables and increases the precision of the estimates. The matching procedure removes approximately 1% employees from our sample, because they belong to strata that do not contain both control and treatment employees.

Although we observe data from many time periods both pre- and post-COVID-19, we aggregate data for each worker into one "pre-treatment" observation and one "post-treatment" observation to address potential autocorrelation. We also cluster standard errors at the level of employee manager, i.e., workers who report to the same manager are in the same "cluster".

Prior work [2] has shown that the extent to which people were exposed to remote work may affect WFH experience. Therefore, we also estimate a version of the DiD model in Eq. 2 that allows the effect of WFH to vary for workers with different levels of prior remote collaboration experience. Specifically, we consider a model with two subgroups of the treatment group, those that had less ($g$ = P-WFH-L) and those that had more ($g$ = P-WFH-M) remote collaboration experience, respectively, and estimate a treatment effect $\delta_L$ and $\delta_M$ for each of the groups.

# 4  Dataset

We measure the causal effect of WFH on a few different collaboration-related metrics from Workplace Analytics [15], a product that aggregates the software usage patterns of Microsoft US employees. The definitions of the metrics are presented in Table 1.

Table 1: Definition of collaboration related metrics from Workplace Analytics [15].

| Metric | Definition |
| --- | --- |
| **Focus hours** | Total number of hours with two or more-hour blocks of time where the person had no meetings. |
| **Email hours** | Total number of hours a person spent sending and receiving emails. |
| **Message hours** | Total number of hours a person spent in instant messages through Microsoft Teams with at least one other person. |
| **Meeting hours** | Total number of hours a person spent in meetings with at least one other person. |
| **Collaboration hours** | Email hours + Message hours + Meeting hours. |

Since the WFH mandate in United States officially went effective in March 2020, for each outcome variable, we computed the average monthly outcome in February and April and treated them as "pre-treatment" and "post-treatment" observations, respectively. We hypothesize that switching to WFH would have a larger effect for employees with less experience in remote collaboration. We measure a worker's remote collaboration experience prior to WFH by computing the percentage of an employee's interactions (including meetings, emails, and instance messages) in January that were with co-workers in different cities or who worked from home. We then assigned workers did less than 20% and more than 70% of their collaborations remotely into low previous remote work experience (P-OFC-L) and high previous remote work experience (P-OFC-M), respectively. Note that the data from January is not used to measure pre- or post-treatment outcomes.

# 5  Findings and results

*WFH lowered collaboration*. As shown in Table 2, total collaboration hours decreased by 1.95% and focus hours increased by 3.57%. Within collaboration, WFH caused a shift from meeting hours, decreasing them by 6.43%, to message hours, increasing them by 27.71%. We do not detect a statistically significant effect of WFH on email hours. The decrease in collaborations can have important implications on how different types of work gets done in remote settings, which we discuss in Section 6.

Table 2: Causal effect of WFH ($\delta/\bar{y}_{\text{feb}}$) for different outcome variables, as well as the amount of change in those outcome variables that is common to both the treatment group and control group ($\lambda/\bar{y}_{\text{feb}}$). The coefficients are normalized by the average level of the outcome variable in the treatment group to give percent differences. We also show the observed change (($\bar{y}_{\text{apr}} - \bar{y}_{\text{feb}}$)/ $\bar{y}_{\text{feb}}$) from February to April.

| % | Collaboration hours | Meeting hours | Message hours | Email hours | Focus hours |
| --- | --- | --- | --- | --- | --- |
| **Effect of WFH ($\delta/\bar{y}_{\text{feb}}$)** | -1.95 * (0.89) | -6.43 *** (1.1) | 27.71 *** (1.86) | 1.29 (1.06) | 3.57 *** (0.71) |
| **Time-fixed effect ($\lambda/\bar{y}_{\text{feb}}$)** | 23.26 *** (0.88) | 30.65 *** (1.09) | 37.20 *** (1.80) | 12.22 *** (1.04) | -6.62 *** (0.71) |
| **Observed** $((\bar{y}_{apr} - \bar{y}_{feb})/\bar{y}_{feb})$ | 21.31 *** | 24.22 *** | 64.90 *** | 13.51 *** | -3.04 *** |

*Contrasting causal estimates with simple observed changes.* The bottom row of Table 2 shows that the observed changes are a combination of the effect of WFH and everything else that changed between February and April, the latter of which is captured by the observed change in outcome for the control group. Looking at these numbers, a simple analysis would suggest that WFH leads to more total collaboration hours, more meeting hours, and fewer focus hours. However, our causal estimates indicate that the effect of WFH was the opposite: WFH decreased collaboration hours, decreased meeting hours, and increased focus hours. In the case of message hours and email hours, a simple comparison suggests that WFH had a much larger effect than our causal estimates indicate. This highlights the importance of causal approaches in understanding the WFH effect.

Table 3 reports the heterogenous effect of WFH ($\delta_M/\bar{y}_{feb-M}$ and $\delta_L/\bar{y}_{feb-L}$) in percentage terms for employees with different levels of remote collaboration experience. As one might expect, the effects are much larger for employees with less remote collaboration experience. The effects for both total collaboration hours and meeting hours are small and statistically insignificant for employees with more remote collaboration experience; for employees with less remote collaboration experience the effects are substantial: -3.1% and -9.23% respectively. For message hours, the effect is statistically significant for employees with more remote collaboration experience, but it is much smaller than for employees with less remote collaboration experience, +5.24% versus +38.5%. For focus hours, there is no effect (insignificant -.75%) for those with more remote collaboration experience and a larger, 4.9%, effect for those with less remote experience. For all variables except email hours, where neither group saw a significant effect, the difference between the two groups is statistically significant.

Table 3: Causal effect of WFH for employees with two different levels of remote collaboration experience: less than 20% of their coworkers were in a different city or working from home (P-OFC-L) and more than 70% of their coworkers were in a different city or working from home (P-OFC-M).

| % | Collaboration hours | Meeting hours | Message hours | Email hours | Focus hours |
|---|---|---|---|---|---|
| **WFH effect for P-OFC-M** $\delta_M/\bar{y}_{feb-M}$ | 1.44 (1.00) | 1.96 (1.28) | 5.24 ** (1.93) | 0.48 (1.14) | -0.75 (0.76) |
| **WFH effect for P-OFC-L** $\delta_L/\bar{y}_{feb-L}$ | -3.10 ** (1.03) | -9.23 *** (1.21) | 38.52 *** (2.28) | 1.81 (1.29) | 4.91 *** (0.82) |
| **Difference** | -4.54 *** (0.59) | -11.19 *** (0.73) | 33.29 *** (1.36) | 1.33 (0.72) | 5.67 *** (0.45) |

The fact that employees with more experience in remote collaboration were less affected by working from home suggests that some of the effect of WFH is due to the need to collaborate remotely as opposed to only the need to physically do one's work at home. The fact that the effects are near-zero for all variables except message hours suggests that physically doing one's work at home is not an important factor for those variables. For message hours, the non-zero effect for those with more remote collaboration experience suggests that for time spent messaging either 1) there is an important effect of physically doing one's work at home instead of in the office, 2) there is a real difference between doing 70% and 100% of one's collaboration remotely, or 3) both. There were not enough employees working in the office with 100% remote collaborators prior to COVID-19 to separate these two effects.

## 6  Implications for the future of remote work

The disruptive effects that many corporations observe during pandemic are mostly likely due to factors other than WFH, so researchers should be cautious about directly generalizing raw changes in work practices in outcomes during the current time to make predictions about WFH in the future. This paper shows the feasibility of using difference-in-differences to make more robust and reliable inferences.

Our study suggests that normal WFH, without a pandemic, could cause an increase in blocks of free time, which people can use to focus on their core tasks. Meanwhile, workers could be collaboratively isolated as meeting time decreases. Given the two effects, the implications for creative work are mixed as such work often requires both deep thought and frequent exchange of ideas. More broadly, since the overwhelming majority of information workers are currently WFH, and post-COVID-19 we expect a new equilibrium with more WFH than the past [1,12,14,18], can companies continue to innovate and create while their workforce is WFH? This is an important longer-term question that the information-work industry should be monitoring.

Our findings also reveal potential opportunities in the design of productivity tools for remote work. (1) As people start to adopt new mediums of communication shifting towards more rapid forms of messaging such as instant messages, more research is needed to understand the special communication needs of remote workers, such as ways to supplement the "water-cooler conversations" that happened in in-person settings. (2) Understanding the heterogeneity of WFH effects with respect to different individual characteristics can lead to the development of tools that provide personalized support, intervention, and recommendation for remote work. This paper shows that people's prior remote collaboration experience is an important dimension of heterogeneity. Our future work will further extend to broader individual attributes.

## 7  Conclusions

We presented a large-scale study on how WFH specifically affects collaboration by leveraging the massive natural experiment happening during COVID-19. Through a difference-in-differences causal inference framework, we show how to control for confounding factors happened concurrently with WFH in order to draw robust and generalizable insights on the effects of WFH. Our findings suggest that the observed changes during the pandemic are mainly due to factors other than WFH, and WFH under normal circumstances is likely to decrease collaboration and increase focus time. Our future work will further expand on our current findings by investigating broader work and productivity measures.